\documentstyle[epsfig]{aipproc}
\begin{document}

\title{The Boson Peak in Amorphous Silica: Results from Molecular
       Dynamics Computer Simulations}

\author{J\"urgen Horbach, Walter Kob, and Kurt Binder}
\address{Institute of Physics, Johannes Gutenberg University,
         Staudinger Weg 7, D--55099 Mainz, Germany}
 
%
\maketitle

\begin{abstract}
We investigate a prominent vibrational feature in amorphous silica, the
so--called boson peak, by means of molecular dynamics computer
simulations. The dynamic structure factor $S(q,\nu)$ in the liquid, as
well as in the glass state, scales roughly with temperature, in
agreement with the harmonic approximation. By varying the size of the
system and the masses of silicon and oxygen we show that the
excitations giving rise to the boson peak are due to the coupling
to transverse acoustic modes.

\end{abstract}

\section{Introduction}
 
In the last few years various scattering techniques, such as neutron,
Raman and X--ray scattering, have been used to investigate the
so--called boson peak, a vibrational feature, which is found in the
frequency spectra of many, typically strong, glass formers at a
frequency of about $1$ THz \cite{boson_peak}. In this context various
mechanisms giving rise to this peak have been proposed, such as certain
localized vibrational modes or scattering of acoustic waves, and also
simple models have been developed that produce an excess over the Debye
behavior in the density of states~\cite{boson_peak_th}.

Especially in the case of silica molecular dynamics computer
simulations have recently been used in order to gain insight into the
nature of the boson peak \cite{taraskin97ab,dellanna97,horbach98}.
Despite the limitations of these simulations, such as the small system
size (of the order of $10^3$--$10^4$ particles) and high cooling rates
(of the order of $10^{12}$ ${\rm K}/{\rm s}$), they are very useful
because they include in principle the full microscopic information in
form of the particle trajectories.  Most of the recent computer
simulation studies have investigated the boson peak within the
harmonic approximation in that the eigenvalues and eigenvectors
of the dynamical matrix have been 
calculated \cite{taraskin97ab,dellanna97}.
In contrast to this method we use the full microscopic information to
determine quantities like the dynamic structure factor $S(q,\nu)$
directly from the particle coordinates. Thus, we are not restricted to
the harmonic approximation and we are able to compare the dynamics of
our silica model in the liquid state with the dynamics in the glass
state. Moreover, by varying parameters like the size of the system and
the mass of the particles we gain information on the character of the
boson peak excitations.

\section{Details of the simulation}

The silica model we use for our simulation is the one proposed by van
Beest {\it et al.} \cite{beest} which is given by
\begin{equation}
  u(r_{ij}) = \frac{q_i q_j e^2}{r_{ij}} + A_{ij} \exp(-B_{ij} r_{ij})
      - \frac{C_{ij}}{r_{ij}^6} \ \ \ {\rm with} \ \ \
       i,j \in \{{\rm Si,O}\}.
\end{equation}
The values of the partial charges $q_i$ and the constants $A_{ij},
B_{ij}$ and $C_{ij}$ can be found in the original publication. The
simulations were done at constant volume keeping the density fixed at
$2.37$ ${\rm g}/{\rm cm}^3$. Our simulation box contains $8016$ particles
with a box length of $48.37$ \AA. We investigate the equilibrium
dynamics of the liquid state as well as the glass state. The lowest
temperature for which we were able to fully equilibrate our system was
$2750$ K. At this temperature we integrated the equations of motion
over $13$ million time steps of $1.6$ fs, thus over a time span of
about $21$~ns. The glass state was produced by starting from two
equilibrium configurations at $T=2900$ K and cooling them to the
temperatures $T=1670$ K, $1050$ K and $300$~K with a cooling rate of
$1.8\cdot10^{12}$ ${\rm K}/{\rm s}$. The details of how we calculated the
time Fourier transformations can be found elsewhere \cite{horbach99}.

\section{Results}
We investigate the high frequency dynamics of silica by means
of the dynamic structure factor
\begin{equation}
S(q,\nu)=N^{-1}\int_{-\infty}^{\infty} dt \exp(i2\pi\nu t)
\sum_{kl} \langle \exp(i{\bf q} \cdot
[{\bf r}_k(t)-{\bf r}_l(0)]) \rangle \quad, \label{sqnue}
\end{equation}
and its self part $S_{{\rm s}}(q,\nu)$ which can be extracted from
Eq.~(\ref{sqnue}) by taking into account only the terms with $k=l$. In
the following we will consider only $S(q,\nu)$ for the oxygen--oxygen
correlations because the oxygen--silicon and the silicon--silicon
correlations behave similarly with respect to the features which are
discussed below.

\begin{figure}[t!]
\centerline{\epsfig{file=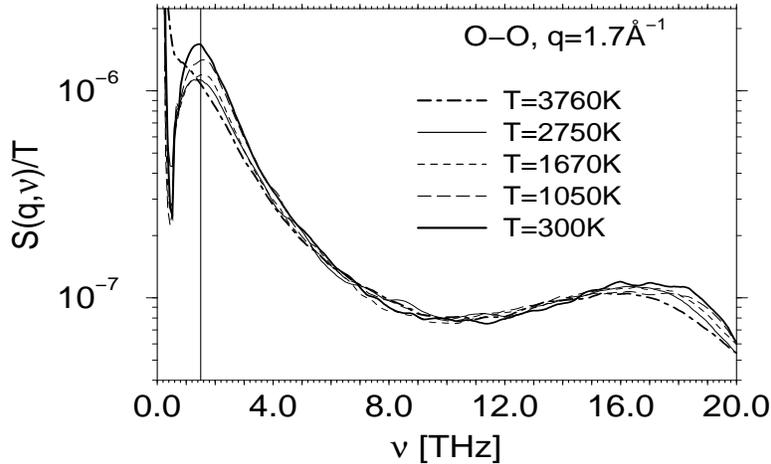,height=2.8in,width=3.8in}}
\vspace{5pt}
\caption{$S(q,\nu)/T$ as a function of frequency for different
temperatures.}
\label{fig1}
\end{figure}
As we have reported elsewhere for the liquid state at the temperature
$T=2900$~K~\cite{horbach98}, apart from optical modes with frequencies
$\nu>20$ THZ, two types of excitations are visible in the dynamic
structure factor for $q>0.23$ \AA$^{-1}$. The first one corresponds to
the boson peak which is located, essentially independent of $q$, around
$1.8$~THz. The second one corresponds to dispersive longitudinal
acoustic modes. Note that the latter are not like longitudinal acoustic
excitations in harmonic crystals because, due to the disorder, they
cannot be described as plane waves. 

Having found the aforementioned two features at $T=2900$ K we will
now look at the temperature dependence of $S(q,\nu)$, which is shown in 
Fig.~\ref{fig1}, by plotting $S(q,\nu)/T$ versus $\nu$
at $q=1.7$ \AA$^{-1}$ in the frequency range below $20$ THz. From this
figure we can conclude that the dynamic structure factor scales roughly
with temperature which is expected if the harmonic approximation is
valid. Moreover we can clearly identify for all temperatures two peaks:
the boson peak located at $1.75$ THz (vertical line) and a peak 
which corresponds to
the longitudinal acoustic modes located around $17$ THz. Even at 
$T=3760$ K
the excitations giving rise to a boson peak at lower temperatures are
at least partially present in that a shoulder can be recognized in the
frequency region of the boson peak.

\begin{figure}[t!]
\centerline{\epsfig{file=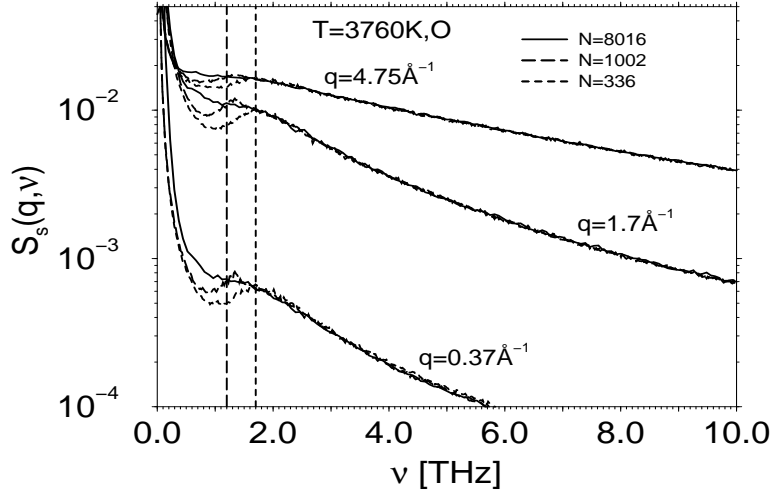,height=2.8in,width=3.8in}}
\vspace{5pt}
\caption{Self part of the dynamic structure factor as a function
of frequency for the system sizes $N=336, 1002$ and $8016$ at the
temperature $T=3760$ K. For the explanation of the vertical lines
see text.}
\label{fig2}
\end{figure}
In order to get some insight into the properties of the vibrational
modes of our silica system at frequencies around $1$ THz we varied the
size of the system at a fixed mass density of $2.37$ ${\rm
g}/{\rm cm}^3$.  Fig.~\ref{fig2} shows the self part of the dynamic
structure factor for $N=336$, $1002$ and $8016$ particles at the
temperature $T=3760$ K and the three $q$ values $0.37$ \AA$^{-1}$,
$1.7$ \AA$^{-1}$ and $4.75$ \AA$^{-1}$.  Whereas the curves for the
different system sizes coincide for frequencies that are larger than a
weakly $N$ dependent frequency $\nu_{{\rm cut}}(N)$, for $\nu <
\nu_{{\rm cut}}(N)$ the amplitude of $S_{{\rm s}}(q,\nu)$ decreases
with decreasing $N$.  Note that $\nu_{{\rm cut}}(N)$ is essentially
independent of the wave--vector $q$. We read off $\nu_{{\rm
cut}}\approx1.7$ THz for $N=336$ and $\nu_{{\rm cut}}\approx1.2$ THz
for $N=1002$. Both frequencies are marked as vertical lines in
Fig.~\ref{fig2}.  $\nu_{{\rm cut}}(N)$ coincides approximately with the
frequency of the transverse acoustic excitation corresponding to the
lowest $q$ value which is determined by the size of the simulation box.
To see that this is the case note
that the lowest $q$ values for $N=336$ and $N=1002$ are $q_{{\rm
min}}=0.37$~\AA$^{-1}$ and $q_{{\rm min}}=0.26$ \AA$^{-1}$,
respectively. In comparison to that the $q$ values we read off from the
transverse acoustic dispersion branch for $T=3760$ K and $N=8016$,
and which correspond to the frequency $\nu_{cut}$ for $N=336$ and 
$N=1002$, are
$0.32$ \AA$^{-1}$ and $0.22$ \AA$^{-1}$, respectively (see \cite{horbach98}). 
That the latter
$q$ values are slightly smaller than the corresponding values for
$q_{{\min}}$ is due to the fact that the transverse dispersion branch
has been determined from the peak maxima $\nu_{{\rm max}}(q)$ in the
transverse current correlation function. So there 
is always a significant contribution of transverse
acoustic modes with frequencies $\nu < \nu_{{\rm max}}(q)$ for a given
$q$. All this can be summarized by saying that the absence of
transverse acoustic modes is connected with a missing of excitations
giving rise to the boson peak.  Therefore, in the smaller systems only
the high frequency part of the boson peak is present.  Moreover, it
seems that the boson peak modes are only fully present, for a given
frequency, if there exist transverse acoustic excitations at the same
frequency.

\begin{figure}[t!]
\centerline{\epsfig{file=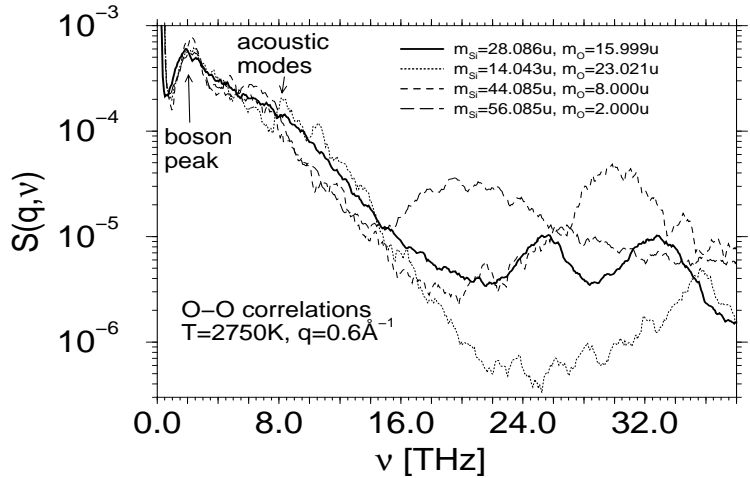,height=2.8in,width=3.8in}}
\vspace{5pt}
\caption{Dynamic structure factor at $T=2750$ K and $q=0.6$ \AA$^{-1}$
under variation of the masses of the silicon and oxygen atoms
such that the mass density is fixed.}
\label{fig3}
\end{figure}
To learn more about the character of the boson peak excitations we
varied also the masses of the silicon and oxygen atoms such that the
mass density remains fixed. Fig.~\ref{fig3} shows $S(q,\nu)$ at $T=2750$
K and $q=0.6$ \AA$^{-1}$ for the four mass pairs $M_1=(28.086,15.999)$,
$M_2=(14.043,23.021)$, $M_3=(44.085,8.000)$, and $M_4=(56.085,2.000)$
where the first and the second number are the masses in atomic units
for silicon and oxygen, respectively. Note that $M_1$ corresponds to
the real masses of silicon and oxygen normally used in our simulation.
From the figure we see that there is a strong dependence on the
mass ratio for the two peaks visible for $M_1$ above $20$ THz which are
due to localized optical modes. In contrast to that, at least within
the accuracy of the statistics of our data, there is no dependence for
the modes giving rise to the boson peak and the acoustic modes which
means that the boson peak excitations cannot be strongly localized. This
supports the aforementioned statement that the boson peak is due to the
coupling to transverse acoustic modes. The fact that the boson peak is
independent of the mass ratio between the silicon and oxygen mass is
also a good test for theoretical models of the boson peak in silica.

In conclusion we investigated the excitations giving rise to the boson
peak by means of molecular dynamics computer simulations.  We find that
the dynamic structure factor $S(q,\nu)$ scales roughly with temperature
in the range $3760$ K $\ge T \ge 300$ K, which means that our
silica system is in this sense quite harmonic even for temperatures as
high as $3760$ K. By calculating $S_{{\rm s}}(q,\nu)$ for different
system sizes we find that the modes contributing to the boson peak are
only fully present at a given frequency if there exist transverse
acoustic modes at the same frequency.  This is supported by the fact
that the height and the width of the boson peak are independent under
the variation of the masses of silicon and oxygen if the mass
density is fixed. So we observe that the boson peak is due to a
coupling to transverse acoustic modes.  Of course, the explanation of
the nature of this coupling is an interesting goal for the future.

\section*{ACKNOWLEDGMENTS}
This work was supported by BMBF Project 03~N~8008~C and by SFB 262/D1
of the Deutsche Forschungsgemeinschaft. We also thank the RUS in 
Stuttgart for a generous grant of computer time on the T3E.

\end{document}